\documentclass[12pt,preprint]{aastex}

\usepackage{xspace}

\newcommand{\g}{{\ensuremath{\mathrm{g}}}\xspace}

\newcommand{\cm}{{\ensuremath{\mathrm{cm}}}\xspace}

\newcommand{\Msun}{{\ensuremath{\mathrm{M}_{\odot}}}\xspace}

\newcommand{\gcc}{{\ensuremath{\g\,\cm^{-3}}}\xspace}

\newcommand{\ltaprx} {\lower .1ex\hbox{\rlap{\raise .6ex\hbox{\hskip .3ex
        {\ifmmode{\scriptscriptstyle <}\else 
                {$\scriptscriptstyle <$}\fi}}}
        \kern -.4ex{\ifmmode{\scriptscriptstyle \sim}\else 
                {$\scriptscriptstyle\sim$}\fi}}}
\newcommand{\gtaprx} {\lower .1ex\hbox{\rlap{\raise .6ex\hbox{\hskip .3ex
        {\ifmmode{\scriptscriptstyle >}\else 
                {$\scriptscriptstyle >$}\fi}}}
        \kern -.4ex{\ifmmode{\scriptscriptstyle \sim}\else 
                {$\scriptscriptstyle\sim$}\fi}}}

\slugcomment{Version \today}

\begin{document}

\title{Nucleosynthesis in Early Supernova Winds II: The Role of Neutrinos}
\author{J. Pruet}
\affil{N Division, Lawrence Livermore National Laboratory, P. O.Box
  808, Livermore, CA 94550}
\email{pruet1@llnl.gov}
\author{R.D. Hoffman}
\affil{N Division, Lawrence Livermore National Laboratory, P. O.Box
  808, Livermore, CA 94550}
\email{rdhoffman@llnl.gov}
\author{S. E. Woosley}
\affil{Department of Astronomy and Astrophysics, UCSC, Santa Cruz, CA,
95064}
\email{woosley@ucolick.org}
\author{H.-T. Janka}
\affil{Max-Planck-Institut f\"ur Astrophysik, Karl-Schwarzschild-Str. 1,
85741 Garching, Germany}
\email{thj@mpa-garching.mpg.de}
\and
\author{R. Buras}
\affil{Max-Planck-Institut f\"ur Astrophysik, Karl-Schwarzschild-Str. 1,
85741 Garching, Germany}
\email{rburas@mpa-garching.mpg.de}

\begin{abstract}

One of the outstanding unsolved riddles of nuclear astrophysics is the
origin of the so called ``p-process'' nuclei from A = 92 to 126. Both
the lighter and heavier $p$-process nuclei are adequately produced in
the neon and oxygen shells of ordinary Type II supernovae, but the
origin of these intermediate isotopes, especially $^{92,94}$Mo and
$^{96,98}$Ru, has long been mysterious. Here we explore the production
of these nuclei in the neutrino-driven wind from a young neutron star.
We consider such early times that the wind still contains a proton
excess because the rates for $\nu_e$ and positron captures on neutrons
are faster than those for the inverse captures on protons.  Following
a suggestion by \citet{fro05}, we also include the possibility that,
in addition to the protons, $\alpha$-particles, and heavy seed, a
small flux of neutrons is maintained by the reaction p($\bar
\nu_e,e^+)$n. This flux of neutrons is critical in bridging the long
waiting points along the path of the $rp$-process by (n,p) and
(n,$\gamma$) reactions.  Using the unmodified ejecta histories from a
recent two-dimensional supernova model by \citet{jan03}, we find
synthesis of $p$-rich nuclei up to $^{102}$Pd.  However, if the
entropy of these ejecta is increased by a factor of two, the synthesis
extends to $^{120}$Te. Still larger increases in entropy, that might
reflect the role of magnetic fields or vibrational energy input
neglected in the hydrodynamical model, result in the production of
numerous $r$-, $s$-, and $p$-process nuclei up to A $\approx$ 170,
even in winds that are proton-rich.

\end{abstract}


\section{INTRODUCTION}

\citet{Bur57} attributed the production of the isotopes heavier than
the iron group to three processes of nucleosynthesis, the $r$- and
$s$-processes of neutron addition, and the $p$-process of proton
addition.  The conditions they specified for the $p$-process, proton
densities $\rho X_p \sim 10^2$ g cm$^{-3}$ and temperatures $T \sim
2 - 3 \times 10^9$ K, were difficult to realize in nature and so other
processes and sites were sought.  \citet{Arn76} and \citet{Woo78}
attributed the production of the $p$-process nuclei to
photodisintegration, a series of ($\gamma,$n), ($\gamma,$p) and
($\gamma,\alpha$) reactions flowing downward through radioactive
proton-rich progenitors from lead to iron. Their ``$\gamma$-process''
operated upon previously existing $s$-process seed in the star to make
the $p$-process, and was thus ``secondary'' in nature (or even
``tertiary'' since the $s$-process itself is secondary). It could only
happen in a star made from the ashes of previous stars that had made
the $s$-process.

Arnould suggested hydrostatic oxygen burning in massive stars as the
site where the necessary conditions were realized; Woosley and Howard,
who discovered the relevant nuclear flows independently, discussed
explosive oxygen and neon burning in a Type II supernova as the likely
site. Over the years, increasingly refined calculations showed that a
portion of the $p$-nuclei could actually be produced as Woosley and
Howard described \citep[e.g.][]{Rau02}. A nagging deficiency of
$p$-process production in the mass range A = 92 - 124 still persisted
though. The production of $^{92,94}$Mo posed a particular problem
since, unless the star had previously experienced a strong
$s$-process, enhancing the abundance of seed above A = 95, there
simply was not enough seed. In massive stars the $s$-process does not
go above mass 90 and so the necessary seed enhancement does not occur.

\citet{hof96} found that large abundances of some $p$-nuclei, and
$^{92}$Mo in particular, could be synthesized in the neutrino-powered
wind blowing from a young neutron star \citep[see
also][]{dun86}. While this wind had chiefly been seen as a way of
making the $r$-process \citep{woo94}, for electron mole numbers, $Y_e
\approx 0.485$, the $p$-nuclei $^{64}$Zn, $^{74}$Se, $^{78}$Kr,
$^{84}$Sr, and $^{92}$Mo were produced in great abundance. It is
important to note in this regard that, while $Y_e = 0.485$ is
nominally neutron rich ($Y_e$ = 0.5 corresponds to neutron, proton
equality), it is still a lot more proton-rich than the $p$-nuclei
themselves (Z/N for $^{92}$Mo = 0.457), so the nucleonic gas contained
some free protons. The $p$-nuclei here were also primary, in the sense
that a star with no initial metallicity would still make the same
composition in its neutrino wind.  There were potential problems,
however, in that the ejection of only a small amount of mass with
$Y_e$ just a little lower than 0.485 resulted in disastrous
overproduction of N = 50 nuclei like $^{88}$Sr, $^{89}$Y, and
$^{90}$Zr. Also the neutron-rich wind failed to produce adequate
amounts of p-process nuclei above A = 92. Though this paper focuses on
early proton-rich outflows, the SN model we study is calculated to
eject a sizable amount of neutron-rich material. It remains to be seen
if very recent simulations predict neutron-rich outflows that satisfy
the conditions needed for efficient synthesis of $^{92}$Mo, or if neutrino
interactions facilitate production of $^{92}{\rm Mo}$ in the neutron rich ejecta 
predicted by these models \citep{ful95}.

Based upon calculations by Jim Wilson, \citet{qia96} pointed out that
$Y_e$ in the wind would naturally evolve though the points necessary
to make these $p$-nuclei and would actually start with a value greater
than 0.5. As other detailed models for core-collapse supernovae became
available, nucleosynthesis was explored in this ``hot, proton-rich
bubble'' by \citet{pru05}, \citet{fro04}, and \citet{fro05}. The latter two
studies found substantial production of nuclei up to A = 84,
including some nuclei traditionally attributed to the $p$-process.  It
seems probable that these winds have also contributed appreciably to
the solar abundances of $^{45}$Sc, $^{49}$Ti and $^{64}$Zn, and,
possibly in a measurable way, to other rare abundances in metal poor
stars.  However, since these same nuclei were already made by other
processes \citep{Woo02,Rau02}, there seemed to be no clear diagnostic
of the proton-rich wind.

Here, following the suggestion of \citet{fro05}, we have revisited
our calculations of the proton-rich wind including, in addition to the
proton captures, the effect of a neutron flux created by p($\bar
\nu_e,e^+)$n. These neutrons have the effect of bridging the
long-lived isotopes along the path of the $rp$-process by (n,p)
reactions and accelerating the flow to heavier elements. For our
standard assumptions regarding expansion rate and neutrino fluxes
\citep{pru05}, we find substantial production of $p$-process
nuclei up to Pd, whereas previously the heaviest major production was
Zn.  If the entropy of the expansion is artificially increased by a
factor of 3 to account for extra energy deposition in the wind
\citep{qia96}, magnetic confinement \citep{Tho03}, or Alfv\'en wave
dissipation \citep{Suz05}, the production of $p$-nuclei extends up 
to A = 170. 

Interestingly, the relevant conditions, $\rho X_p \approx 10^3$ g
cm$^{-3}$ and T = $2 \times 10^9$ K resemble those originally proposed
for the $p$-process by B$^2$FH. Key differences, however, are that all
the species produced here are primary and the process occurs on a
shorter time scale - just a few seconds - owing to the ``effective'' acceleration of
weak decays by (n,p) reactions.

\section{Nucleosynthesis in an exploding 15$\,$M$_{\odot}$ Star}

Our fiducial model for exploring nucleosynthesis in the proton-rich
wind is the explosion of a 15$\,$M$_{\odot}$ star 
\citep{jan03,bur05}. An earlier paper \citep[][henceforth Paper I]{pru05}
studied nucleosynthesis in this same model but did not account for the
influence of neutrino interactions. The present study
includes charged-current capture on free nucleons \citep{mcful95,mcful96}
and neutral-current
spallation of nucleons from alpha particles \citep{woo90}.  Many other details of
the supernova model and associated nuclear-network calculations
relevant to this work can be found in Paper I.

The ejecta of the deepest layers of the supernova can be divided into
two categories - hot bubble ejecta and winds. Material in the hot
bubble originates from a region outside the neutron star that is
driven convectively unstable by neutrino heating. This material does
not have to escape the deep gravitational well found at the neutron
star surface. As a result, modest conditions characterize these
outflows: $s/k_b \sim 20-30$ and $Y_e\lesssim 0.52$.  Winds originate from
the surface of the neutron star and are pushed outwards along gentle
pressure gradients caused by neutrino heating
\citep{dun86,qia96}. These outflows have relatively high entropies
($s/k_b \sim 50-80$), high electron fractions ($Y_e$ as large as 0.57), and
short initial expansion timescales.  Tables 2 \& 3 in Paper I provide
a brief summary of initial conditions in different bubble and wind 
trajectories.

In the absence of neutrinos, very little synthesis occurs in the early
proton-rich outflows for nuclei heavier than A= 64. Compared to
observed solar abundances, proton-rich winds that are not subject to a
neutrino fluence are copious producers of $^{45}$Sc and $^{46,49}{\rm
Ti}$. Bubble trajectories, on the other hand, tend to favor the
synthesis of $^{64}{\rm Zn}$, $^{46,49}{\rm Ti}$ and some Co or Ni
isotopes. The nuclear flow stops at $^{64}$Ge
for two reasons: the long weak decay lifetime (with respect to the
expansion timescale of the neutrino winds) of this and other even-Z
even-N 'waiting point' nuclei and the small Q values for proton
capture on these nuclei.

\subsection{Estimating the late time evolution}

The simulations of \cite{jan03} followed the 
explosion of the supernova in 2 dimensions until about 450 ms after core bounce. At
this time, typical temperatures in the bubble trajectories were around
4 billion degrees. At the end of the 2D simulation the SN was mapped onto a 1D grid. 
The subsequent evolution of the SN, including winds emitted by the neutron star, was
followed until about 1300 ms after core bounce. At this last time typical asymptotic temperatures
in the early wind were just over 2 billion degrees. 

Since most of the
interesting nucleosynthesis occurs for $T_9 \leq 2$, it is necessary to
extrapolate outflow conditions to later times. Details of the
extrapolation were described in Paper I which considered two
estimates. In the first, the expansion already calculated is assumed
to continue homologously with no deceleration. This gives a useful
lower bound on the expansion timescale. However, in reality the
outflow will quickly catch up to the outgoing shock. It is more
reasonable then to assume that escaping matter enters a phase where it
moves with the shock speed. Conditions for this can be estimated from
1D supernova simulations.

For the present paper we focus on the more realistic extrapolation
that mimics the late time slowing of the wind.  The temperature and
density evolution was extrapolated as in Paper I. Specifically, the
trajectories found by \cite{jan03} were smoothly merged with those
calculated for the inner zone of the same $15\Msun$ star by
\cite{Woo95}.  To avoid discontinuities in the entropy at the time
where the two calculations were matched the density in the previous 1D
calculation was changed to match that in the trajectories found by
Janka et al.  Our procedure likely leads to an underestimate of the
expansion timescale at late times because the explosion energy and
shock velocity in the calculations of Woosley \& Weaver were somewhat
larger than in the more recent 2D simulations. 

In order to include the influence of the neutrinos from the cooling
neutron star additional assumptions are needed. For the neutrino
temperatures and luminosities we used the same values calculated at
approximately 1 second in the simulation of \cite{jan03}. These values
are shown in table \ref{nuTable} and are assumed to remain
constant. This may be a questionable assumption - neutrino spectra
actually harden and their luminosities fall as the neutron star
shrinks.  However, it is estimated that uncertainties in the
extrapolation of trajectories to low temperatures are greater than
uncertainties arising from our simple treatment of the neutrino
spectral evolution.

To obtain the integrated neutrino exposure seen by outflowing
material, it is also necessary to make assumptions about the evolution
of the radial velocity. In all cases, it was assumed that the radial
velocity at times greater than those followed in the simulations was
constant at $v_r=4\cdot 10^{8}{\rm cm \,sec^{-1}}$. This is close to the
velocity of the outgoing shock in the 2D calculation of this fairly
low energy explosion. Again, there may be some inconsistency between
our treatment of the late time expansion and our adopted asymptotic
radial velocity. A more sophisticated approach might scale the
expansion time estimated from the calculations of Woosley \& Weaver to
reflect the relatively small shock velocity found in the 2D
simulations. It is shown in section \ref{modest} that our calculations
for nucleosynthesis are relatively insensitive to changes in the
late-time outflow velocity.

\subsection{Results}

Figure \ref{allFig} shows results of our calculations that include
neutrino captures for production factors characterizing
nucleosynthesis in proton-rich outflows leaving the early SN. The
production factor for an isotope $i$ is defined here as
\begin{equation}
\label{proddef}
P(i)=\sum_j{M_j \over M^{\mathrm{ej}}}{X_j(i) \over X_{\odot,i}}.
\end{equation}
In this equation, the sum is over all trajectories, $X_j(i)$ is the
mass fraction of nuclide $i$ in the $j$th trajectory, $X_{\odot,i}$ is
the mass fraction of nuclide $i$ in the sun \citep{lod03}, $M_j$ is
the mass in the $j$th trajectory, and $M^{\mathrm{ej}}=13.5\,\Msun$ is
the total mass ejected in the SN explosion. 

The lower panel in figure \ref{allFig} shows production factors
characterizing nucleosynthesis in just the hot bubble. This is
comprised of the 40 trajectories described in Table 1 of Paper I.  By
comparing with Table 5 from Paper I we see that nucleosynthesis in the
bubble material is not greatly changed when neutrino captures are
included. This is because the bubble material is far from the neutron
star and experiences few neutrino captures relative to the number of
seed nuclei in this low entropy environment.  The story is much
different for the early wind shown in the middle panel of figure
\ref{allFig}. This wind is comprised of the 6 trajectories defined in
Table 2 of Paper I.  In these outflows, neutrinos convert free protons
into neutrons. Capture of these neutrons by (n,p) reactions on
long-lived nuclei along the path of the $rp$-process accelerate the
progress to nuclei as heavy as Pd. Final conditions for these wind
trajectories are provided in table \ref{5512570_final}. We defer a
discussion of the nuclear flows and the potential for production of
heavier elements to section \ref{broaderStudy}.

\begin{figure}
\includegraphics[scale=0.6,angle=270]{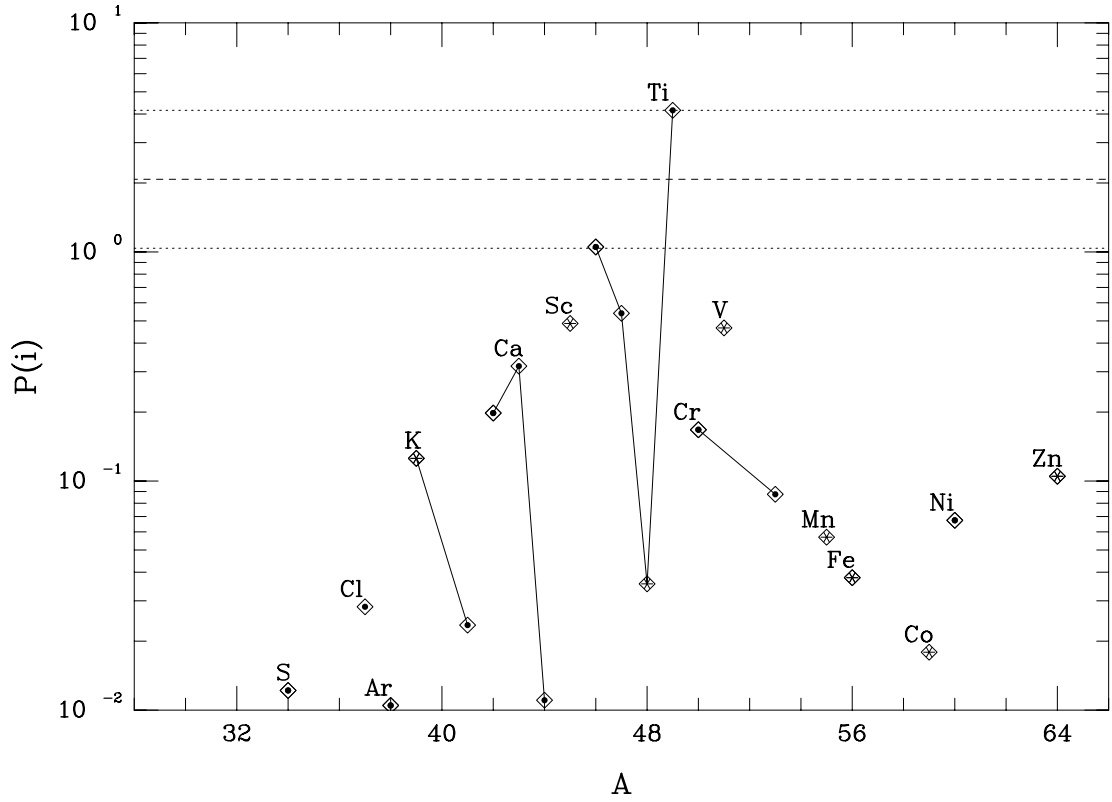}
\caption{Influence of neglecting neutrino captures on nucleosynthesis
in the early wind. This figure shows integrated production factors for
the six wind trajectories given by the fiducial supernova
simulation. Apart from the neglect of neutrinos, the trajectories
studied here are the same ones represented by the middle panel of the
previous figure.
\label{allFigNoNeutrinos}}
\end{figure}

For comparison, we also show, in figure \ref{allFigNoNeutrinos}, the
integrated results (summed over the same six wind trajectories) of
calculations that do {\sl not} include neutrino capture on
protons. Apart from the neglect of neutrinos the trajectories studied
here are identical to those described in the middle panel of figure
\ref{allFig}.  The differences are dramatic. When neutrino captures
are ignored nucleosynthesis stops some 40 mass units lower than when
they are included.

\subsection{Influence of Modest Changes to the Early Wind and Neutron Star}
\label{modest}

The calculations in the previous section are quite uncertain owing
to both an uncertainty in the supernova explosion model and to our procedure
for extrapolating expansion after the explosion calculation has
stopped.  Modest alterations are reasonable to the asymptotic wind
velocity, the electron fraction of the wind, and neutrino capture
rates in the outflow.  More extreme changes that might reflect the
influence of novel physical processes operating in the early wind are
explored in section \ref{broaderStudy}.

\begin{itemize}
\item{Influence of a Larger Asymptotic Wind Velocity

The supernova studied by Janka et al. has a relatively low kinetic
energy at infinity, 0.6$\times 10^{51}$ erg. More energetic
explosions would give rise to shock velocities larger than the $4\cdot
10^8{\rm cm \,sec^{-1}}$ adopted here.  To estimate nucleosynthesis in more
energetic supernovae, we show in the first column of table \ref{modestTable}
production factors for nuclei synthesized in a wind characterized by
an asymptotic velocity of $10^9{\rm cm \,sec^{-1}}$. This is more typical of supernovae
with late time kinetic energies near $10^{51}$ erg. Apart from the
increase in asymptotic velocity all other properties of the 6
trajectories comprising the early wind were kept unchanged from those
found by simulation. 

By comparing with the middle panel of figure \ref{allFig}, one sees
that changing the asymptotic outflow velocity has little effect on
nucleosynthesis.  This may be partly an artifact of our definition of
``asymptotic'' as applying only to times after the last point given by
the supernova simulation. To test this we also ran simulations where
the outflow velocity was set to $10^9{\rm cm \,sec^{-1}}$ once the
temperature of the wind fell below 2.5 billion degrees.  Again the
differences were small.  }

\item{Effect of 5$\%$ changes to $Y_e$

Uncertainty in the neutrino spectral evolution or the dynamics of the
wind near the neutron star could affect $Y_e$ in the wind.  The second
and third columns of table \ref{modestTable} show the influence of
changing the electron fraction up or down by $5\%$. This change was
applied to all 6 trajectories comprising the early wind. Other
characteristics of these trajectories were left unchanged.

It is evident that relatively small changes to the electron fraction
can have a big impact on nucleosynthesis. If $Y_e$ is decreased by
5$\%$ production of $p$-isotopes near mass 100 is lost and replaced by
modest synthesis of some proton-rich Kr and Sr isotopes. A 5$\%$
increase in $Y_e$ leads to large production of Pd and Cd isotopes. The
reason for the large impact of changing $Y_e$ can be understood in
terms of the number of protons available for capturing neutrinos.  The
most proton-rich trajectory found by simulations had $Y_e=0.570$. This
corresponds to a mass fraction of free protons $X_p\approx
2Y_e-1=0.140$. Increasing $Y_e$ by five percent corresponds to a 40
percent increase in $X_p$.}

\item{Changes to the Neutrino Capture Rates

The last two columns of table \ref{modestTable} show the influence of
halving and doubling the luminosity of electron neutrinos and
anti-neutrinos. Such large changes to the luminosities are probably
unlikely, but might reflect plausible uncertainties in the local
neutrino capture rate experienced by the wind. For example, the wind
material might first catch up with the outgoing shock at a larger
radius than found by simulations. Our neglect of the temporal
evolution and finer spectral details of the neutrinos might also
result in modest changes to the capture rates.

From table \ref{modestTable} it is seen that halving or doubling the
charged current neutrino captures rates is roughly equivalent to
decreasing or increasing $Y_e$ by 5$\%$.  }

\end{itemize}

\subsection{Implications}

Galactic chemical evolution studies indicate that production factors
in the whole star for isotopes exclusively produced in core-collapse
supernovae must be of order 10 \citep{mat92}.  As noted before
\citep{pru05,fro05}, this implies that the current simulations predict
a hot bubble ejecta that could explain the origin of $^{46,49}{\rm
Ti}$ and $^{64}{\rm Zn}$. Implications for the early wind are more
interesting. Without any tuning or rescaling of wind conditions, the
simulations of \cite{jan03} predict a wind that efficiently
synthesizes several interesting $p$-nuclei - including the elusive
isotopes $^{96,98}{\rm Ru}$ and $^{102}{\rm Pd}$ - in
near solar relative proportions. Overall these predict about 5-10
times too much yield of the most proton-rich stable Ru and Pd
isotopes.  In the previous section it was shown that small and
plausible changes to the electron fraction can alleviate this
overproduction.

\section{Nucleosynthesis in More Extreme Conditions}
\label{broaderStudy}

In this section we consider nucleosynthesis in outflows for which
neutrino, proton, and neutron-induced reactions on heavy seed can
produce still heavier nuclei. This may occur because the proton to
seed ratio is higher - as happens if the entropy is higher - or the
production of neutrons by p($\bar \nu,e^+)$n is greater. Reasons why
the entropy might be higher are discussed in the conclusions.

\subsection{Basic considerations}

Qualitatively, the nucleosynthesis we are describing occurs in three,
or possibly four stages. First, in the outgoing wind, all neutrons
combine with protons leaving an excess of unbound protons - much like
in the Big Bang. As this combination of alphas and protons cools below
$5 \times 10^9$ K, a small fraction of the alphas recombine to produce
nuclei in and slightly above the iron group - $^{56}$Ni, $^{60}$Zn,
and $^{64}$Ge. Flow beyond $^{64}$Ge is inhibited however by strong
reverse flows, especially (p,$\alpha$) reactions. 

The second stage occurs as the temperature declines below $\sim 3
\times 10^9$ K. A combination of (p,$\gamma)$ and (n,p) reactions
carries the flow, still close to the Z = N line, to heavier nuclei.
For A$<$92 the flow in the present calculations passes through the
even-even N=Z nuclei. After $N=Z=44$ ($^{88}$Ru) the character of the
flow changes.  Effective synthesis of the next even-even nucleus
($^{92}$Pd) is prevented in part by the small proton separation energy
of $^{91}$Rh - the proton capture parent of $^{92}$Pd.  As in the
analogous $r$-process, just how far the flow goes in a particular
trajectory depends on the proton-to-seed ratio and especially on the
number of neutrons per seed produced by p($\bar \nu,e^+)$n. All
interesting nuclei in this stage are made as proton-rich progenitors.

The third stage occurs as the temperature drops below $1.5 \times
10^9$ K and charged-particle reactions freeze out. Neutrons are still
being produced by p($\bar \nu,e^+)$n, however, albeit at a reduced
rate (both because the neutrino luminosity declines and the distance
to the neutron star is getting greater). (n,p) reactions now drive
material towards the valley of beta stability. Because the nuclei
involved are unstable to positron decay anyway, this only
accelerates the inevitable. The atomic mass, A, does not change. It should
be noted, however, that just as the $r$-process can 
contribute to nuclei made by the $s$-process that are unshielded against
$\beta$ decay of more neutron-rich isobars, so too can the $\nu-rp$ process
considered here contribute to $s$-nuclei that are unshielded 
on the proton-rich side. That is, in addition to nuclei that are 
designated as ``$r,s$'', there may also be nuclei one should consider as
``$p,s$''. 

The fourth stage only occurs in the most extreme situation where the
number of neutrons produced by neutrinos is quite large compared with
the number of seed nuclei. Then (n,p) reactions not only carry the
flow at low temperature back to the valley of beta stability, but
(n,$\gamma$) reactions carry it beyond - {\sl to the neutron rich side
of the periodic chart}, even in the presence of a large abundance of
free protons. This is a novel version of the $r$-process that actually
works best when the {\sl proton} abundance is large but the temperature too
low for proton addition. The protons are just a source of neutrons.

\subsection{A Basic Parameter of the Process - $\Delta_n$}

The most interesting part of the nucleosynthesis occurs during the
later stages of the outflow as the material cools. In the absence of 
an important neutrino flux the final isotopic yields are determined by
an interplay between $({\rm p},\gamma)$, $(\gamma,{\rm p})$ and
$\beta^+$ processes as well as details of how these reactions fall out
of equilibrium.  When neutrino capture on free protons is important,
nuclei are pushed to higher isospin and mass via ${\rm (n,p)}$ and
(n,$\gamma$) reactions.

As a first approximation neutrinos will be important if they create an appreciable number
of neutrons per heavy nucleus. The ratio of created free neutrons to heavy nuclei
is 
\begin{equation}
\label{yneq}
\Delta_n=\frac{X_pn_{\nu}}{X_{\rm heavy}/\bar{A}}\approx 60\frac{(2Y_e-1)n_{\nu}}{1-X_{\alpha}-X_p},
\end{equation}
where $X_{\rm heavy}$ is the mass fraction of elements heavier than $\alpha$ particles
and $\bar{A}$ is an effective average atomic number. In eq.~\ref{yneq} 
\begin{equation}
n_{\nu}=\int{\lambda_{\bar{\nu}} dt},
\end{equation}
is the net number of neutrinos captured per free proton at temperatures smaller than about $3\cdot 10^9$K. 
Here 
\begin{equation}
\label{lambdanu}
\lambda_{\bar{\nu}} \approx 0.06 \cdot
\frac{L_{\bar{\nu}_e}}{10^{52}{\rm erg/sec}}
\frac{T_{\bar{\nu}_e}}{4{\, \rm MeV}} \left( \frac{10^8{\rm cm}}{r}\right)^2
\end{equation}
is the rate at which each free proton captures $\bar{\nu}_e$'s \citep{qia96}.
In eq.~\ref{lambdanu} $L_{\bar{\nu}_e}$ is the luminosity
of electron anti-neutrinos, $T_{\bar{\nu}_e}$ is an effective temperature for these neutrinos and $r$ is the radius of
the material from the neutrino sphere.

To estimate the relation between $\Delta_n$ and nucleosynthesis
consider a mass element of unit volume co-moving with the wind. The
number of free protons in this mass element is
\begin{equation}
n_p=X_p \rho N_A.
\end{equation}
These free protons are destroyed by anti-neutrino capture
at a rate 
\begin{equation}
\dot{n}_p =-n_p \lambda_{\bar{\nu}}.
\end{equation}
Neutrons created when anti-neutrinos capture onto protons are subsequently
absorbed by nuclei. The evolution of the free neutron abundance is
then set by a competition between neutrino and nuclear processes
\begin{equation}
\label{xndot}
\dot{X}_n=X_p \lambda_{\bar{\nu}} - \rho X_n\sum_i\frac{\lambda_i X_i}{A_i}.
\end{equation}
Here the sum is over all isotopes $i$, $A_i$ and $\lambda_i$ represent the atomic number and
rate at which species $i$ absorbs free
neutrons. As a matter of convention $\lambda_i$ here represents 
the rate at which a single atom of species $i$ absorbs
neutrons when the free neutron density is one mol per unit volume. If we introduce 
an average neutron absorption rate $\bar{\lambda}$ the
destruction rate appearing on the right hand side of eq.~\ref{xndot} can be written
\begin{equation}
\sum \frac{\lambda_i X_i}{A_i}=X_{\rm heavy} \frac{\bar{\lambda}}{\bar{A}}.
\end{equation}

Neutrons are principally absorbed in (n,p) and $({\rm n},\gamma)$ reactions, with 
$(n,\alpha)$ reactions playing a smaller role (due to the larger coulomb barrier 
in the exit channel). Table \ref{repRates} shows rates
for a sample of nuclei found in proton-rich outflows. 
At 2 billion 
degrees typical values of $\lambda_i$ for even-even proton-drip line 
nuclei (those bordering the line separating the proton-bound nuclei from the proton-unbound
nuclei) with mass near A$=$72 are around $10^8-10^9{\rm cm^3/mol\cdot sec}$. 
This implies 
a very rapid neutron destruction rate. For example, at a typical density of $10^4$ g cm$^{-3}$
and a mass fraction of heavy nuclei equal to $10^{-3}$, a neutron is absorbed in less than a
microsecond. Since this is much shorter than the material expansion rate it is fair to 
treat the neutron abundance as being in equilibrium 
\begin{equation}
X_n\approx \frac{X_p\lambda_{\bar{\nu}} \bar{A}}{\rho X_{\rm heavy} \bar{\lambda}}.
\end{equation}
An estimate for the abundance of free neutrons also gives an estimate for the destruction rate
of an atom of species $i$:
\begin{equation}
\dot{X}_i=-X_iX_n\lambda_i \rho \equiv -\tilde{\lambda}_i X_i.
\end{equation}
Here we have defined an effective destruction rate that reflects the competition between different nuclear species
for scarce neutrons
\begin{equation}
\tilde{\lambda}_i=\left(\frac{100}{\rm sec}\right)\left( \frac{X_p\bar{A}}{10} \right) \left( \frac{10^{-2}}{X_{\rm heavy}} \right)
\left( \frac{\lambda_{\bar{\nu}}}{0.1{\rm \,sec^{-1}}} \right) \left( \frac{\lambda_i}{\bar{\lambda}} \right),
\end{equation}
which is enormous. This equation says for plausible outflow conditions a given species can be entirely 
destroyed by neutrino-produced neutrons in times as short as $\sim 10$ ms. 
It also implies a very small equilibrium neutron abundance: $X_N\lesssim 10^{-12}$ for $\rho\approx 10^4$ g cm$^{-3}$.

One of the most interesting questions relates to how high in mass the
nucleosynthesis will proceed. For the purpose of making first
estimates we can suppose the starting inventory of nuclei to be
concentrated near mass 60. We will also suppose that $\lambda_i$ is
independent of species for nuclei with $A\geq 60$. In this case the
number of neutrons captured by a heavy nucleus is just $\Delta_n$. A more careful
treatment of $\lambda_i$ is not presented since results from detailed 
network-based calculations are given in the next section. With the
above assumptions the mass fraction of species $j$ is
\begin{equation}
\label{howHigh}
\frac{X_j}{X_{\rm heavy}} \approx \frac{\Delta_n^j}{j!}e^{-\Delta_n};\,\,\,j=N-30.
\end{equation}
Here we have defined a species to include all nuclei of a given isotone. 
Depending on just how fast the effective absorption rates
$\tilde{\lambda}_i$ are one might or might not suppose that decay of
nuclei with odd-N is dominated by weak processes. This is because odd-N drip-line nuclei
can have rather fast $\beta^+$ rates (see table \ref{repRates}).
When the neutron absorption rates are slower than 
these weak rates  
one would take $j\approx (N-30)/2$ in the above equation. 

Though eq.~\ref{howHigh} is crude it can be used as a rough guide to gauge
the influence of free neutrons created from neutrino captures. As an
example, suppose one neutron is created per heavy nucleus
($\Delta_n=1$). In this case eq.~\ref{howHigh} suggests that the
relative mass fraction of nuclei that have captured a single neutron
is about $1/e$, while the relative mass fraction of nuclei that
have captured 4 neutrons is about 20 times smaller. If we use a factor
of ten decrease in mass fraction as a rough cutoff, this implies that
an appreciable abundance of nuclei with mass up to A$\approx
60+4\cdot4\approx 76$ will be synthesized. Here we have supposed that
a unit increase in neutron number is accompanied by a unit increase in
proton number. Neutron capture on odd-N proton-drip line nuclei has
been neglected since the $\beta^+$ rates of these nuclei are much
faster than the assumed neutron capture rate of $~\sim1/$sec.  As
another example, suppose that $\Delta_n=5$. In this case the relative
mass fraction of nuclei that have captured 5 neutrons is about
20/$e^5$, while the relative mass fraction of nuclei that
have captured 11 neutrons is about 20 times smaller.  This suggests
appreciable synthesis of nuclei up to mass $A\approx 60+4\cdot
11\approx 100$.  Here we have again assumed that the change in atomic
number is twice the change in neutron number and that weak processes
alone destroy odd-N nuclei. These assumptions probably lead to an
overestimate for the increase in $A$ in this case because a neutron
capture rate of $\sim 5/$sec is comparable to the weak rates of many
proton drip-line nuclei. 

The above considerations suggest a first constraint on conditions synthesizing 
$p$-nuclei with mass near 130
\begin{equation}
\Delta_n\gtrsim 10.
\end{equation}
A second constraint comes from considering the evolution of the
outflow at low temperatures.  As $T$ falls below about 1.5$\cdot10^9$K
charged-particle capture rates begin to freeze out. Neutrons, on the
other hand, are still rapidly absorbed.  These neutron captures push
the flow toward stability and away from progenitors of $p$-nuclei. If
low-temperature neutrino-induced neutron production is significant,
even the very neutron-rich $r$-process nuclei are synthesized. Minimal
destruction of $p$-nuclei implies a second constraint
\begin{equation}
\label{con2}
\Delta_n(T\lesssim 1.5\cdot 10^9) < {\rm \,a\,\,few}.
\end{equation}
Conversely, efficient synthesis of $r$-process isotopes in these
proton-rich outflows requires the production of several neutrons per
heavy nucleus at low temperatures.

\subsection{Results of Network Calculations}

Outflows characterized by the production of many free neutrons per
heavy nucleus (large $\Delta_n$) can be realized in different
ways. For example, the timescale characterizing the expansion of the
outflow around the time that $\alpha$-particles are synthesized might
be small, the flow might be held close to the neutron star for an
extended period, or $Y_e$ might be very large.  For simplicity we
consider implications of changing the entropy of the outflow.  Apart
from the influence of neutrino captures occurring at low temperature,
the precise mechanism by which $\Delta_n$ is increased is not so
important for nucleosynthesis.  Changes to the entropy of the hot
bubble are not considered. Neutrino capture is not very pronounced in
this portion of the outflow. As well, the hot bubble material does not
begin close to the neutron star, so it is hard to see how an
appreciable increase in the entropy of this material could be
achieved.

Figure \ref{2sfig} shows the influence of doubling the entropy in the
early wind.  Like the middle panel of figure \ref{allFig}, this figure
gives integrated production factors for a wind comprised of six
trajectories.  Each of these trajectories has twice the entropy, but
is otherwise identical to, a counterpart trajectory from the
simulation. For definiteness the
increase in entropy was assumed to influence only the density evolution.
The evolution of temperature with time was assumed to be the same as that
found from simulation. To a fair approximation doubling or tripling of the
entropy corresponds to dividing the density by a factor of two or
three.

Doubling the entropy in the early wind results in values of $\Delta_n$
ranging from about 0.6 to 10.  The increased number of neutron
captures results in efficient synthesis of nuclei as heavy as mass
125. By contrast, for all of the unmodified wind trajectories 
$\Delta_n$ is less than about 3 and efficient synthesis stops
around Ru. 

Figure \ref{3sfig} shows the influence of tripling the entropy in the
early wind.  Again - this shows integrated production factors for six
trajectories that each have larger entropy, but that are otherwise
identical to, unaltered trajectories described by the middle panel of
figure \ref{allFig}. These modified high entropy wind trajectories
have values of $\Delta_n$ ranging from about 1 to 22. This results in
efficient synthesis of nuclei as heavy as mass 170. In other words,
increasing the number of neutrons captured per heavy nucleus by 10
pushes the flow some 40 units higher in mass.

\subsection{Influence of Neutrino Capture at Low Temperatures}

\begin{figure}
\epsscale{0.7}
\plotone{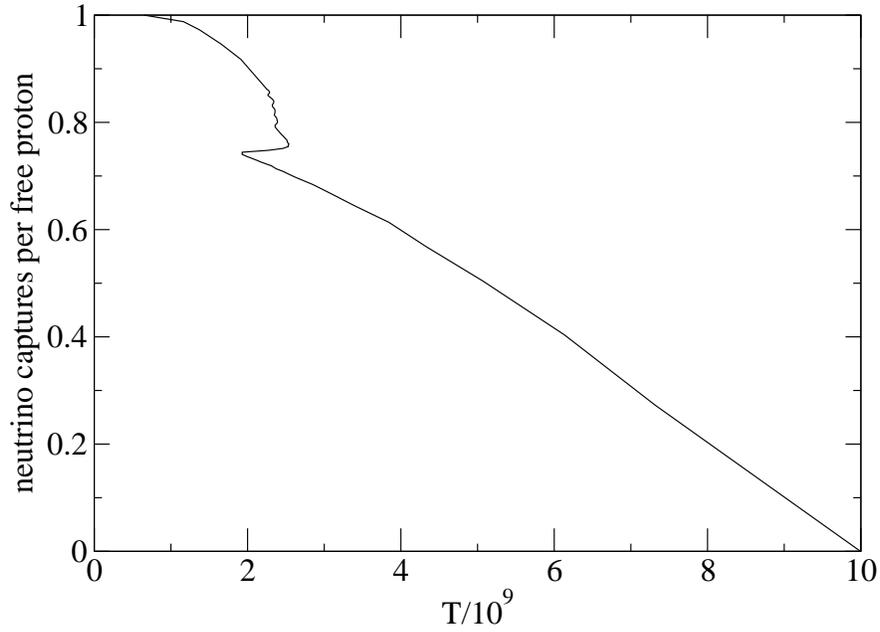}
\caption{Evolution of the number of neutrino captures per free proton
as a function of temperature in a wind trajectory calculated by
\cite{jan03}. This trajectory is characterized by $s/k_b\approx 77$
and $Y_e=0.57$. The y axis has been normalized to unity. Note that
only a small fraction of the neutrino captures occur at low
temperature.
\label{ev1}}
\end{figure}

Though outflows predicted by simulations have values of $\Delta_n$
that are too small to allow synthesis of $A\sim 130$ $p$-nuclei, they
naturally satisfy the constraint (eq.~\ref{con2}) on the relative
number of neutrino captures occurring while the wind material is so
cold that charged particle captures have frozen out. Figure \ref{ev1}
shows the evolution of $\int{dt/r^2}$ as a function of temperature for
the wind outflow characterized by $s/k_b\approx 77$ and $Y_e=0.57$. It is
seen that the fraction of neutrino captures occurring at low
temperatures smaller than 1.5 billion degrees is quite small, less
than about $5\%$.

To illustrate the influence of neutrino captures occurring while the
outflow is cold enough that charged particle reactions have frozen
out we modified the entropy doubled version of trajectory 6 (Table
\ref{5512570_final}) to be held close to the neutron star at
temperatures less than 1.5 billion degrees.  A relatively modest
modification in the outflow results in the capture of about 5 neutrons
per heavy nucleus at temperatures lower than $1.5\cdot 10^9$K.  The
first four columns of Table \ref{slowcompare} shows a comparison
between nucleosynthesis in this slow outflow and nucleosynthesis in an
outflow with the nominal radial velocity of $4\cdot 10^8{\rm cm \,sec^{-1}}$.
Since the only difference between these two trajectories is the
evolution of radius with time at low temperatures, all differences in
nucleosynthesis arise from late-time neutron production.  It is seen
that a couple of neutrons produced at the wrong time can be
detrimental to the synthesis of some $p$-process isotopes.

We also show in Table \ref{slowcompare} the influence of a great
number of neutrons produced at low temperatures. Again, this was
studied by modifying just the radial profile at temperatures less than
1.5 billion degrees of the entropy doubled version of trajectory 6.
The last two columns of the table show nucleosynthesis in this
trajectory in which about 20 free neutrons are created per heavy
nucleus at low temperatures. Most of the isotopes shown are
$r$-process isotopes. It is perhaps remarkable that the 2nd and
possibly 3rd $r$-process peak elements can be synthesized in these
proton-rich environments. It may be difficult, however, to have ejecta
that are both cold enough and close enough to the neutron star to 
experience the necessary neutrino irradiation.

\subsection{Details of the Nuclear Flows}
\label{detailflows}
In all trajectories studies, regardless of initial electron fraction
or entropy, nucleosynthesis begins with $^{12}$C produced early on by
the reaction sequence
$\alpha$($\alpha$n,$\gamma$)$^{9}$Be($\alpha$,n)$^{12}$C.  By the time
T$_9\sim 3$ the iron group has already been assembled. Strong
$(\alpha,\gamma )$ and pairs of (p,$\gamma$) and ($\alpha$,p)
reactions continue to populate the even-Z even-N $\alpha$-nuclei up to
$^{56}$Ni and $^{60}$Zn.  The flow mostly travels along the Z=N line
and does not stray more than two neutrons from it for any element up
to zinc.  This continues until the charged-particle reactions freeze
out (T$_9\sim 1.5$).

\begin{figure}
\begin{center}
\includegraphics[angle=270,width=\columnwidth]{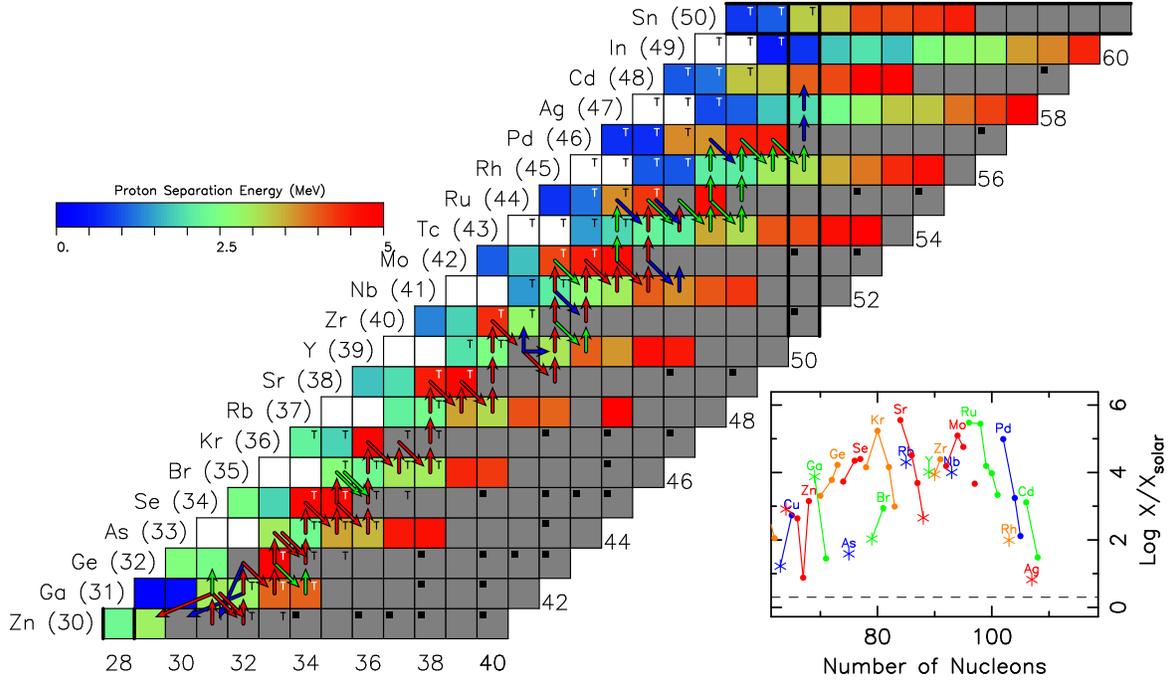}
\caption{Net nuclear flows in the (Z,N)-plane from zinc to tin when material in the
unmodified wind outflow of trajectory 6 has a temperature T$_9=2.05$
and density $\rho=2.7\times10^4\gcc$.  The net nuclear flow (in units
of sec$^{-1}$) is defined as the product of abundance, density, and
reaction rate in the forward (charge or mass increasing) direction
minus a similar quantity for the inverse reaction. Strong and
electromagnetic flows begin at the center of a target nucleus and end
as an arrow in the product nucleus.  Any flow that starts off-center
represents weak decay.  Net nuclear flows are plotted in three
strengths: red (strong), green (intermediate) and blue (weak), with
values that are between a factor of 1.0 to 0.1, 0.1 to 0.02, and 0.02
to 0.01 of the value of the largest flow in the figure, respectively.
The largest flow here is $^{61}$Zn(p,$\gamma$)$^{62}$Ga ($1.75\times 10^{-4}$ sec$^{-1}$).
Stable species are
represented by a filled black square in the upper left corner. Each
nucleus is color coded according to the legend by the value of its proton
separation energy. Proton unbound nuclei are colored white. Nuclei
with $S_p>5$ MeV are colored gray. A "T" is plotted in the upper right-hand
corner for nuclei whose binding energy was extrapolated from measured
masses \citep{Aud95}.  Production factors at
the time shown are given in the inset (the stable isotopes depicted
include the abundances of all radioactive progenitors that will
eventually decay to them).  As discussed in the text the classical
$rp-$process waiting points ($^{64}$Ge, $^{68}$Se, $^{72}$Kr, and $^{76}$Sr) are
bypassed by (n,p) reactions.
\label{j570_s1_psp_zn2sn}}
\end{center}
\end{figure}

Characteristics of the nucleosynthesis at lower temperatures depends
sensitively on the influence of neutrino captures. To illustrate the
influence of p($\bar \nu_e,e^+)$n reactions we begin with a discussion
of nucleosynthesis in trajectory 6, which is characterized by the weak
production of a few neutrons per heavy nucleus.  Important nuclear
flows occurring when material in this trajectory has a temperature
$T=2.05\times10^9$K are shown in figure \ref{j570_s1_psp_zn2sn}. It is seen that the
dominant flows (red arrows) are due to proton-capture (p,$\gamma$) reactions.
These can proceed until a proton unbound (denoted by a white square) or small (blue) $S_p$
energy is encountered. Unlike the $rp$-process, here we have a neutron
abundance and though small it allows (n,p) or $\beta^+$
reactions to populate the next lowest isobar. The (p,$\gamma$) flow is
governed by the separation energies.

The end result for this trajectory is the production of the light 
$p$-process nuclei from Kr to Pd.
The (n,p) reactions can continue to carry the flow even at low
temperatures because such reactions on targets a few neutrons to the
proton side of stability typically have positive Q-values (i.e. no
thresholds). The flow to heavier nuclei eventually stops when the
charged particle reactions freeze out (T$_9 \leq 1.5$) and at late
times (once the waiting points are passed) when (n,p) and (n,$\gamma$) reactions or
weak decay direct the flows toward stability.

An interesting although unfortunate occurrence is the low (relative to
Ru and Pd) production of the most abundant $p$-nucleus in nature,
$^{92}$Mo. As the proton capture flow moves up the N=46 isotones (see
figure \ref{j570_s1_psp_zn2sn}) it is stopped in part because
$^{91}$Rh has a small proton separation energy. This prevents
efficient population of $^{92}{\rm Pd}$ (Z=N=46) and so breaks the
pattern of synthesizing the even-even N=Z nuclei.  As a consequence the
flow detours towards stability until reaching N=47 and N=48. The
result is that the radioactive progenitor for $^{92}$Mo is now the
odd-odd nucleus $^{92}$Rh.  The flow moves very quickly through this
nucleus (as well as through $^{92}$Ru), and little is left for
decay at the end of this trajectory.

It is notable that the heavier $p$-nuclei, $^{96,98}$Ru and
$^{102}$Pd, are co-produced in amounts that might explain their solar
abundances. Their radioactive progenitors are associated with nuclei
in the two nearby closed shells.  Heavier $p-$ nuclei ($^{102}$Cd
etc.) are not made here because the flow failed to populate isotopes
in the Z=50 proton shell.  $^{92}$Mo is the only one of these
intermediate $p$-nuclei that (for now) appears to have an odd-Z
progenitor.  We note however that the flow goes through regions where
the possible error on $S_p$ is potentially large (indicated by the
'T', meaning the value of $S_p$ was from an extrapolation from
measured values). More accurate measurements here would be most
welcome, and would have significant impact on our understanding of
$p$-process nuclei and their solar abundances since this material is
ejected (unlike the case in x-ray bursts). As an example, the
uncertainty in the proton separation energy of $^{91}{\rm Ru}$ is
about 600 keV. A plausible 1 MeV increase in this separation energy
results in a 50$\%$ increase in the yield of $^{92}$Mo in trajectory
6.

\begin{figure}
\begin{center}
\includegraphics[angle=270,width=\columnwidth]{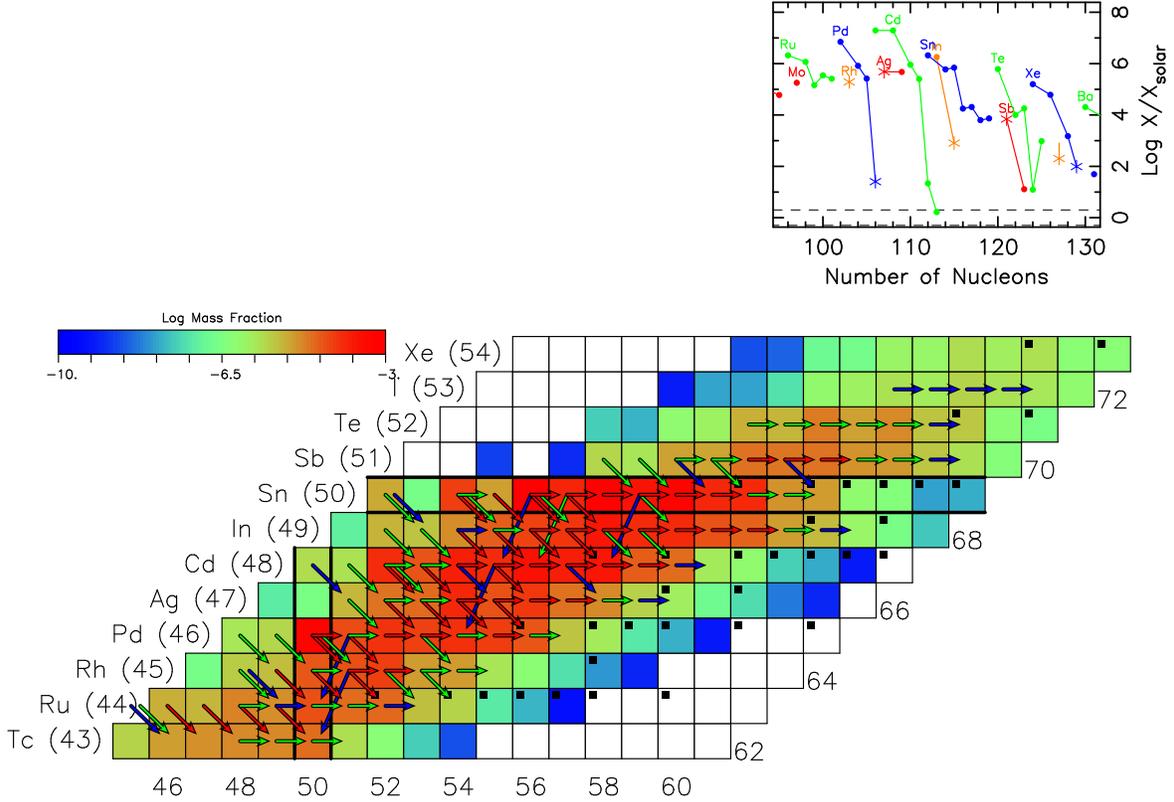}
\caption{Net nuclear flows from technetium to xenon in the modified
(entropy doubled) wind trajectory with $Y_e=0.570$.  At the time shown
here T$_9=1.01$, $\rho=1.2\times 10^3$, and $Y_e=0.561$. The reactant mass
fractions are X(p)=0.122, X(n)=$10^{-12}$, and X($\alpha$)=0.865. The
largest abundance is X($^{108}$Sn)$=7.1\times 10^{-4}$, the largest
flow depicted is $^{109}$Sn(n,$\gamma$)$^{110}$Sn ($7.91\times
10^{-7}$ sec$^{-1}$).  The charged particle reactions have frozen out,
leaving (n,p) and (n,$\gamma$) reactions to carry the flow rapidly
towards stability (before the onset of weak decay). This allows
$p-$nuclei like $^{112}$Sn and $^{120}$Te to be made as themselves via
neutron capture reactions.
\label{j570_s2_tc2xe}}
\end{center}
\end{figure}

Figure \ref{j570_s2_tc2xe} shows nucleosynthesis in a trajectory
characterized by the production of many free neutrons per heavy
nucleus. This trajectory has an initial electron fraction $Y_e=0.570$ and
entropy $s/k_b\approx 148$. It was constructed by doubling the entropy
in the 2-D simulation of trajectory 6 (Table \ref{5512570_final}).
For this modified outflow $\Delta_n\approx 22$.

When material in this modified trajectory reaches a temperature of
about T$_9 \sim 2$, (p,$\gamma$) reactions on $^{110}$Sn (with $S_p
\ge 5$ MeV) pierce the Z=50 closed proton shell.  At the time shown in
figure \ref{j570_s2_tc2xe} (2.21 sec, T$_9=1.01$), the charged
particle reactions have frozen out, but the flow has entered an area
where weak decay has yet to dominate. Instead, (n,p) and (n,$\gamma$)
reactions carry the flow rapidly toward stability. The $p$-nuclei of
Ru, Pd, and Cd are all made as radioactive progenitors in the closed
neutron (Ru) and proton (Pd \& Cd) shells.  We are in a very novel
regime, where one can synthesize $p$-nuclei (like $^{112}$Sn and
$^{120}$Te) via neutron capture reactions.

\section{Conclusions}

Current supernova simulations, without modification, provide the
necessary conditions required to explain the origin of a number of
$p$-process isotopes between A = 92 and 126 whose origin in nature has
always been unclear.  The site is the proton-rich bubble that powers
the explosion and the early neutrino-powered wind that develops right
behind it. The synthesis is primary, so a neutron star derived from a
metal poor progenitor star would produce the same yields (so long as
the neutron star itself had the same properties).  Very metal
deficient stars formed from these ejecta would be characterized by a
excess of both $p$-process nuclei and $r$-process nuclei compared to
the $s$-process, but since there is no element that is dominantly
$p$-process, observational diagnostics may be difficult.

\begin{figure}
\epsscale{0.6}
\plotone{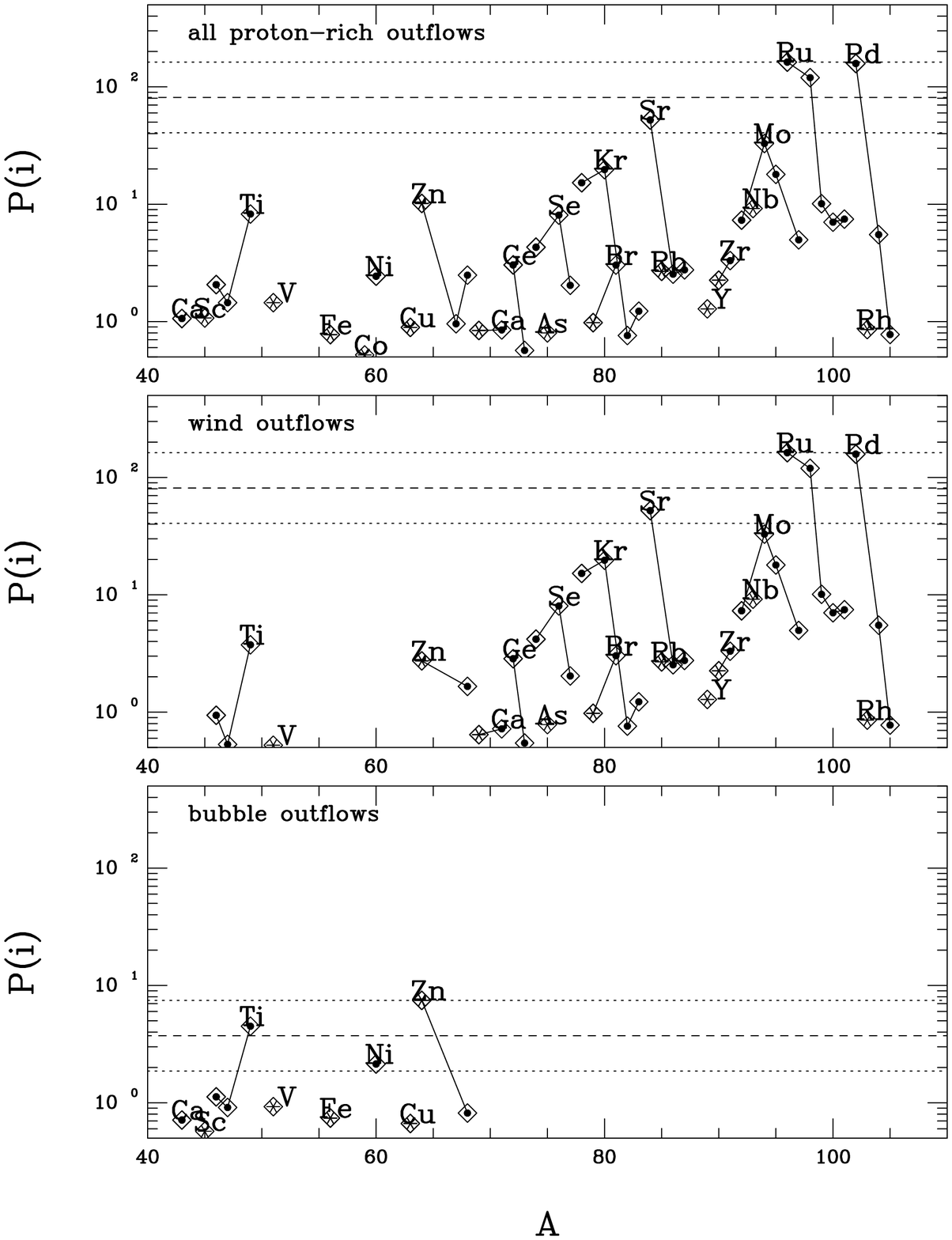}
\caption{Production factors characterizing nucleosynthesis in
proton-rich trajectories occurring during the explosion of the 15
solar mass star studied by \cite{jan03}. The lowest panel gives the
contribution of just the proton-rich hot bubble trajectories to the
net production factors. These trajectories are characterized by the
weak production of few neutrons per heavy nucleus.  The number of
neutrons created by weak interactions per heavy nucleus ($\Delta_n$)
in these flows spans a range from about zero to 0.05.  The middle
panel gives the contribution of just the proton-rich wind trajectories
to the total nucleosynthesis. These winds are characterized by
relatively high entropies and electron fractions, and reach low
temperatures near the neutron star.  Neutrinos result in the
production of between about 0.2 and 3.2 free neutrons per heavy
nucleus in the wind trajectories.  The top panel shows net production
in all proton-rich outflows. This is the sum of the lower two
panels. In each panel solid lines connect isotopes of a given element.
The most abundant isotope for a given element is indicated with a
star.  Any isotope surrounded by a diamond indicates it was made
chiefly as a radioactive progenitor.  We draw horizontal lines at the
production factor value of the largest nucleus produced, and a factor
of two and four less than that respectively.
\label{allFig}}
\end{figure}

\begin{figure}
\epsscale{0.7}
\includegraphics[scale=0.6,angle=270]{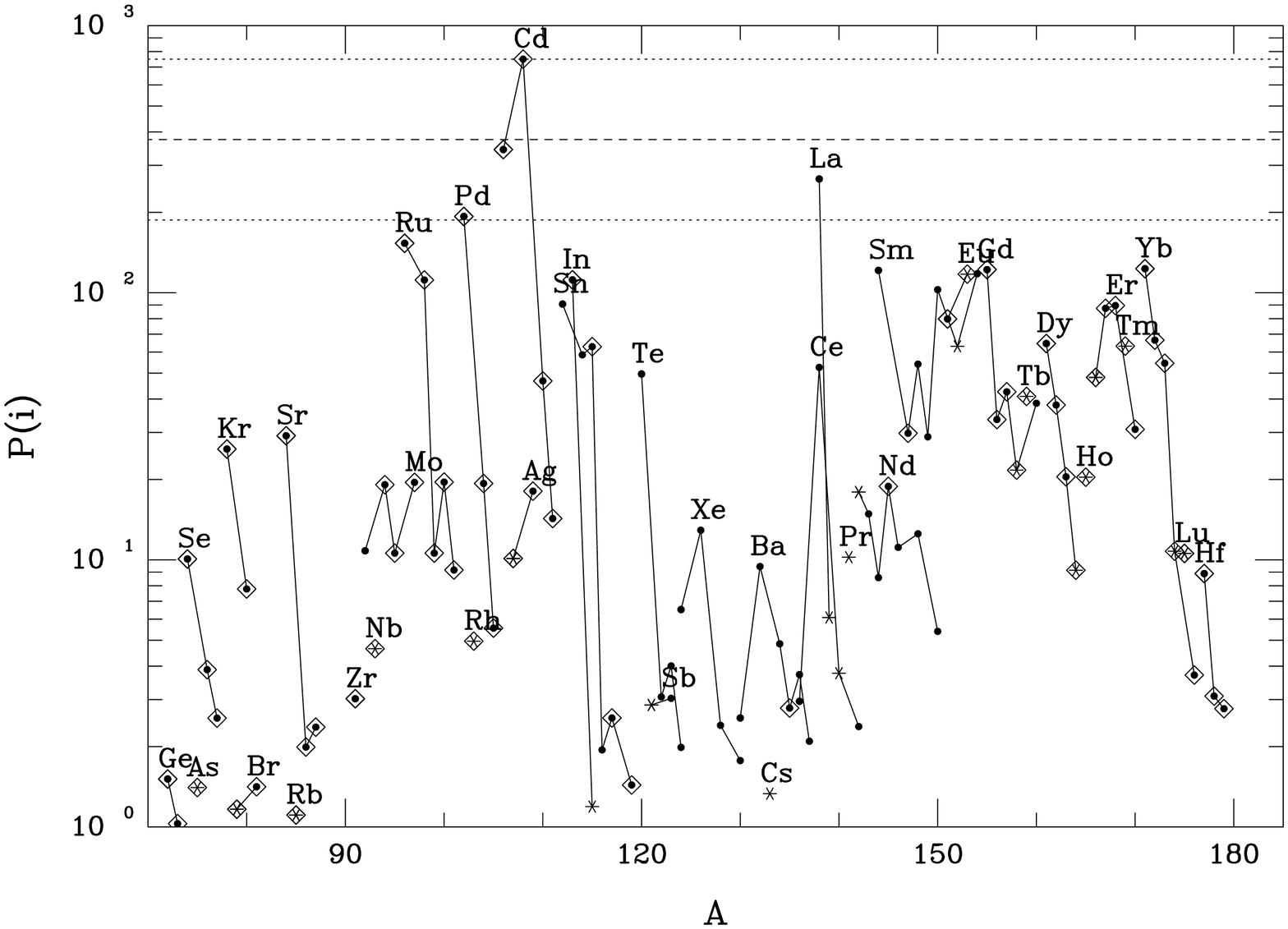}
\caption{Integrated production factors calculated under the assumption
that all six outflows comprising the early wind have three times the
entropy found by \cite{jan03}.  Apart from the change in entropy these
outflows are assumed to have the same mass, electron fraction, and
evolution of radius and temperature with time as the outflows
represented in the middle panel of figure \ref{allFig}. For these very
high entropy outflows $\Delta_n$ spans the range from 1 to 22.
\label{3sfig}
}
\end{figure}

In particular, large quantities of $^{96,98}{\rm Ru}$ and $^{106}$Pd
are produced in our calculations (Fig. \ref{allFig}).  Synthesis of
$p$-process isotopes as heavy as $^{120}{\rm Te}$ can also be achieved
by only factor-of-two modifications to the entropy of the baseline
simulation.  It is interesting in this regard to note that an even
larger increase in entropy is needed later in the {\sl neutron}-rich
wind for the efficient synthesis of the $r$-process isotopes
\cite[e.g.,][]{qia96}. This is quite possibly informing us of some
additional heating mechanism that operates in the mass outflow during
the first few seconds of a neutron star's life. Possible mechanisms
are magnetic field entrainment of the outflowing matter \citep{Tho03},
magnetic energy dissipation \citep{qia96}, acoustic energy input
\citep{qia96,Bur05}, or Alfv\'en wave dampening \citep{Suz05}. None of
these were included in the present supernova model, but we varied the
entropy to determine qualitatively their effect.

In the more extreme, but still physically reasonable case that the
entropy is multiplied by three, the synthesis extends all the way to 
$^{168}$Yb, with the accompanying production of many isotopes normally
attributed to the $s$-process and even the $r$-process. 

Somewhat disappointingly, none of our calculations produce 
a large overabundance of $^{92}$Mo compared to surrounding isotopes
(though some do make 10\% of the necessary value). This may reflect 
either the fact that $^{92}$Mo has another origin, e.g., the same neutrino-powered
wind a few seconds later when $Y_e$ = 0.485, or uncertainties in the 
nuclear physics. In the current study, the $^{92}$Mo that is made is
produced as the odd-odd progenitor $^{92}$Rh. This does not take advantage 
of the extra stability that would be afforded by an even-even nucleus 
like $^{92}$Pd, let alone the magic neutron shell of $^{92}$Mo itself. 
Indeed the binding energies and lifetimes of nuclei in the vicinity
of $^{92}$Pd are quite uncertain. 

\begin{figure}
\includegraphics[scale=0.6,angle=270]{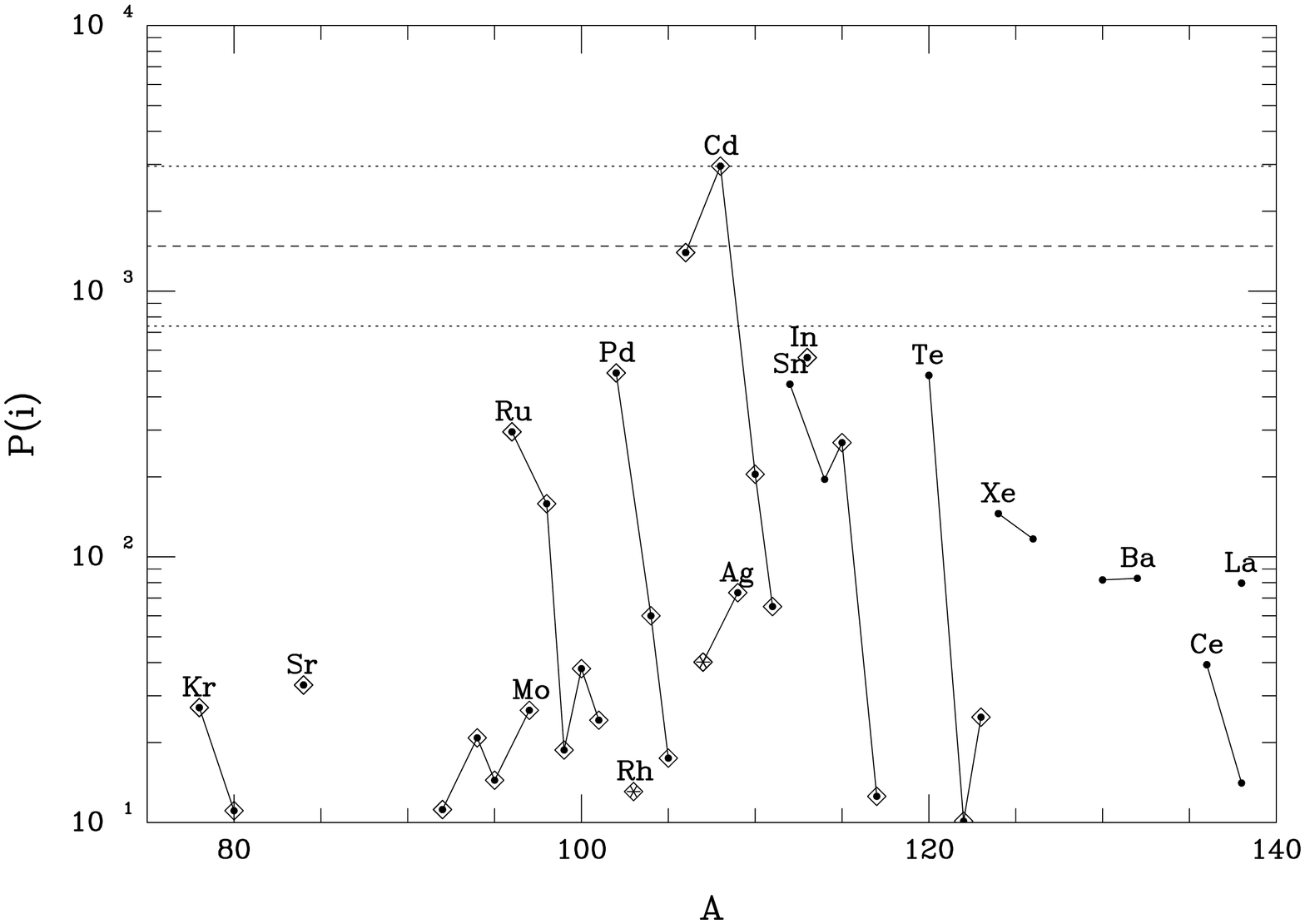}
\caption{Integrated production factors calculated under the assumption that all six
outflows comprising the early wind have twice the entropy found by \cite{jan03}. 
Apart from the change in entropy these outflows are assumed to have the same mass,
electron fraction, and evolution of radius and temperature with time as
the outflows represented in the middle panel of figure \ref{allFig}. The increase in entropy 
results in fewer seed nuclei and values of $\Delta_n$ ranging from about 0.6 to 10.6. 
\label{2sfig}
}
\end{figure}

An important aspect of the synthesis calculated here is that
none of the $p$-nuclei are made as themselves; all have proton-rich
progenitors. Many of these progenitors are so unstable that even their
masses and lifetimes are not measured, let alone their cross sections
for interacting with neutrons and protons. A similar situation is
encountered in the $rp$-process in Type I x-ray bursts
\citep[e.g.][]{Sch01}, a critical difference being that the isotopes
made here are actually ejected and contribute to the solar inventory
of heavy elements. The study of these is a major goal for nuclear
astrophysics experiments of the future, like the Rare Isotope
Accelerator (http://www.anl.gov/ria/index.html). For now, we can only
note that these nuclear uncertainties are almost certainly responsible
for a large fraction of the spread in production factors in,
e.g. Figs. \ref{allFig} and \ref{2sfig}.

This study has explored only a relatively limited set of outflow
parameter space based upon simple modifications to trajectories found
in one particular simulation.  Further studies will surely be carried
out by us and others, but we have identified a key physical parameter,
$\Delta_n$ (eq. 2), which characterizes the solution for various
combinations of time scale, $Y_e$, and entropy.  $\Delta_n$ is
essentially a dimensionless measure of the number of neutrons produced
by neutrino capture on protons compared to the number of heavy seed
nuclei. Surveys on a finer grid of $\Delta_n$ than were used here
will be interesting.

This work was performed under the auspices of the U.S. Department
of Energy by the University of California Lawrence Livermore National
Laboratory under contract W-7405-ENG-48.
It was also supported, in part, by the SciDAC Program of
the US Department of Energy (DC-FC02-01ER41176), the National Science
Foundation (AST 02-06111), and NASA (NAG5-12036) and, in Germany, by
the Research Center for Astroparticle Physics (SFB 375) and the
Transregional Collaborative Research Center for Gravitational Wave
Astronomy (SFB-Transregio 7).

\clearpage

\clearpage

\clearpage

\clearpage

\clearpage

\clearpage

\clearpage

\clearpage
\begin{deluxetable}{cccc}
\tabletypesize{\scriptsize}
\tablecaption{Parameters characterizing the different neutrino species at 925 ms
post core bounce in the the simulation of \cite{jan03}. 
\label{nuTable}}
\tablewidth{0pt}
\tablehead{
\colhead{Species}
& \colhead{$T_{\nu}(MeV)$} 
& \colhead{$L_{\nu}/10^{51}{\rm erg/sec}$}}
\startdata
$\nu_e$ & 3.86    & 17.9 \\
$\bar{\nu}_e$ & 4.62  & 17.7\\
$\nu_x\tablenotemark{a}$ & 4.9  & 24.5
\enddata
\tablenotetext{a}{Represents any of the $\mu$ and $\tau$ neutrinos and antineutrinos.}
\end{deluxetable}

\clearpage
\begin{deluxetable}{ccccccccc}
\tabletypesize{\scriptsize}
\tablecaption{Final conditions for early wind outflows\tablenotemark{a}
\label{5512570_final}}
\tablewidth{0pt}
\tablehead{
\colhead{traj.}
& \colhead{$Y_e$}
& \colhead{$s/k_b$}
& \colhead{$X(p)$}
& \colhead{$X(\alpha)$}
& \colhead{$X_{H}$}
& \colhead{$X(^{56}{\rm Ni}$)}
& \colhead{\%\tablenotemark{b}}
& \colhead{$\Delta_n$\tablenotemark{c}}}
\startdata
 1 & 0.539 & 54.8  & 0.078 & 0.614 & 0.307 & 0.244 & 80 & 0.2 \\
 2 & 0.548 & 58.0  & 0.095 & 0.714 & 0.190 & 0.135 & 71 & 0.4 \\
 3 & 0.551 & 76.7  & 0.101 & 0.822 & 0.075 & 0.043 & 57 & 1.7 \\
 4 & 0.551 & 71.0  & 0.102 & 0.796 & 0.101 & 0.063 & 62 & 1.1 \\
 5 & 0.556 & 74.9  & 0.113 & 0.831 & 0.054 & 0.025 & 46 & 2.9 \\
 6 & 0.558 & 76.9  & 0.115 & 0.840 & 0.043 & 0.014 & 33 & 3.2 \\ 
\enddata
\tablenotetext{a}{At the end of the simulations T$_9\approx 0.65$.}
\tablenotetext{b}{The percentage of heavy nuclei that was $^{56}{\rm Ni}$}
\tablenotetext{c}{Estimate from eq.~\ref{yneq} of the number of neutrons created by neutrino capture at temperatures less than 3 billion degrees.}
\end{deluxetable}

\clearpage
\begin{deluxetable}{cccccccccc}
\tabletypesize{\scriptsize}
\tablecaption{Influence of Modest Changes to the Early Wind and Neutron Star on Nucleosynthesis
\label{modestTable}}
\tablewidth{0pt}
\tablehead{
\colhead{isotope\tablenotemark{a}} 
& \colhead{$\log(P)$\tablenotemark{a}} 
& \colhead{isotope\tablenotemark{b}} 
& \colhead{$\log(P)$\tablenotemark{b}}
& \colhead{isotope\tablenotemark{c}} 
& \colhead{$\log(P)$\tablenotemark{c}}
& \colhead{isotope\tablenotemark{d}} 
& \colhead{$\log(P)$\tablenotemark{d}}
& \colhead{isotope\tablenotemark{e}} 
& \colhead{$\log(P)$\tablenotemark{e}}}
\startdata
$^{98}$Ru & 2.09 &   $^{84}$Sr & 1.22 & $^{102}$Pd & 3.14 & $^{84}$Sr & 1.25 & $^{102}$Pd & 3.07 \\
$^{102}$Pd & 2.06 & $^{80}$Kr & 0.93 & $^{106}$Cd & 2.72 & $^{80}$Kr & 0.91 & $^{96}$Ru & 2.90 \\
$^{96}$Ru & 1.86 & $^{78}$Kr & 0.92 & $^{96}$Ru & 2.69 & $^{96}$Ru & 0.82 & $^{106}$Cd & 2.71 \\
$^{84}$Sr & 1.72 & $^{49}$Ti & 0.74 & $^{98}$Ru & 2.65 & $^{98}$Ru & 0.71 & $^{98}$Ru & 2.66 \\
$^{94}$Mo & 1.36 & $^{76}$Se & 0.69 & $^{108}$Cd & 2.10 & $^{78}$Kr & 0.65 & $^{108}$Cd & 2.14 \\
$^{80}$Kr & 1.20 & $^{96}$Ru & 0.56 & $^{104}$Pd & 1.98 & $^{49}$Ti & 0.59 & $^{84}$Sr & 1.94 \\
$^{95}$Mo & 0.99 & $^{74}$Se & 0.39 & $^{84}$Sr & 1.79 & $^{76}$Se & 0.56 & $^{104}$Pd & 1.92 \\
$^{93}$Nb & 0.89 & $^{64}$Zn & 0.34 & $^{101}$Ru & 1.76 & $^{94}$Mo & 0.55 & $^{94}$Mo & 1.86 \\
$^{106}$Cd & 0.87 & $^{94}$Mo & 0.30 & $^{94}$Mo & 1.70 & $^{102}$Pd & 0.39 & $^{101}$Ru & 1.82 \\
$^{99}$Ru & 0.83 & $^{72}$Ge & 0.23 & $^{99}$Ru & 1.61 & $^{64}$Zn & 0.26 & $^{100}$Ru & 1.77 \\
\enddata
\tablenotetext{a}{\,Largest production factors in a wind with an asymptotic velocity of $10^9{\rm cm \,sec^{-1}}$.}
\tablenotetext{b}{\,Largest production factors in a wind with $Y_e$ decreased by 5$\%$ relative to the 
value found by simulation. Apart from the change in $Y_e$ for each of the 6 wind trajectories all other characteristics
are the same as found by simulation.}
\tablenotetext{c}{\,Largest production factors in a wind with $Y_e$ increased by 5$\%$ relative to the 
value found by simulation. Apart from the change in $Y_e$ for each of the 6 wind trajectories all other characteristics
are the same as found by simulation.}
\tablenotetext{d}{\,Largest production factors in a wind where the charged current neutrino capture rates 
are half those found by simulation. Other characteristics of the wind were left unchanged.}
\tablenotetext{e}{\,Largest production factors in a wind where the charged current neutrino capture rates 
are twice those found by simulation. Other characteristics of the wind were left unchanged.}
\end{deluxetable}

\clearpage
\begin{deluxetable}{cccc}
\tabletypesize{\scriptsize}
\tablecaption{Neutron absorption and weak decay rates for some important proton-rich nuclei\tablenotemark{a}.
\label{repRates}}
\tablewidth{0pt}
\tablehead{
  \colhead{parent nucleus} 
& \colhead{$\lambda({\rm n,p})$\tablenotemark{b}}
& \colhead{$\lambda({\rm n,\gamma})$\tablenotemark{b}}
& \colhead{$\lambda$($\beta^+ +$e.c.)\tablenotemark{c}}}
\startdata
$^{64}{\rm Ge}$ & 6.4$\cdot 10^8$ & 4.5$\cdot 10^5$ & 0.01 \\
$^{66}{\rm As}$ & 7.7$\cdot 10^8$ & 9.6$\cdot 10^5$ & 7.24 \\
$^{68}{\rm Se}$ & 7.6$\cdot 10^8$ & 1.1$\cdot 10^6$ & 0.02 \\
$^{70}{\rm Br}$ & 1.0$\cdot 10^9$ & 2.0$\cdot 10^6$ & 8.67 \\
$^{80}{\rm Zr}$ & 1.6$\cdot 10^9$ & 4.5$\cdot 10^6$ & 0.18 \\
$^{116}{\rm Cd}$& 3.3$\cdot 10^{-13}$ & 1.7$\cdot 10^7$ & 0.\\
\enddata
\tablenotetext{a}{Taken from the statistical model calculations of \cite{Rau00}.}
\tablenotetext{b}{Stellar Rate (${\rm cm^3}/{\rm mol\cdot sec}$) at T$_9=2$, $\rho=1$ g/cc.} 
\tablenotetext{c}{Weak rate in units of sec$^{-1}$.}
\end{deluxetable}

\clearpage
\begin{deluxetable}{ccccccc}
\tabletypesize{\scriptsize}
\tablecaption{Influence of the weak-production of neutrons at low temperatures
on nucleosynthesis.\label{slowcompare}}
\tablewidth{0pt}
\tablehead{
\colhead{isotope\tablenotemark{a}} 
& \colhead{$\log(X/X_{\odot})$\tablenotemark{a}} 
& \colhead{isotope\tablenotemark{b}} 
& \colhead{$\log(X/X_{\odot})$\tablenotemark{b}}
& \colhead{isotope\tablenotemark{c}} 
& \colhead{$\log(X/X_{\odot})$\tablenotemark{c}}}
\startdata
$^{108}$Cd & 7.33 & $^{114}$Sn & 7.20 &  $^{122}$Sn & 6.45      \\
$^{106}$Cd & 6.99 & $^{113}$In & 7.08 &   $^{124}$Sn & 6.22     \\
$^{120}$Te & 6.87 & $^{112}$Sn & 7.00 &    $^{116}$Cd & 6.13    \\
$^{113}$In & 6.82 & $^{108}$Cd & 6.87 &    $^{110}$Pd & 6.01    \\
$^{112}$Sn & 6.75 & $^{115}$Sn & 6.85 &   $^{123}$Sb & 6.01     \\
$^{102}$Pd & 6.50 & $^{120}$Te & 6.55 &  $^{105}$Pd & 5.89      \\
$^{115}$Sn & 6.48 & $^{102}$Pd & 6.43 &  $^{103}$Rh & 5.87      \\
$^{124}$Xe & 6.41 & $^{106}$Cd & 6.34 &  $^{104}$Ru & 5.80    \\
$^{114}$Sn & 6.37 & $^{98}$Ru & 6.25  &   $^{111}$Cd & 5.76     \\
$^{110}$Cd & 6.31 & $^{126}$Xe & 6.00 &   $^{121}$Sb & 5.71    
\enddata
\tablenotetext{a}{\,Largest overproduction factors for an
outflow characterized by the weak-production of very few neutrons per heavy nucleus at temperatures lower
than $2\cdot 10^9$K.}
\tablenotetext{b}{\,Largest overproduction factors for an outflow characterized by the production
of about 5 neutrons per heavy nucleus at temperatures lower than $1.5\cdot 10^9$K. Apart from having a smaller
radius at low temperatures this trajectory is identical to that represented by the first two
columns of this table.}
\tablenotetext{c}{\,Largest overproduction factors for an outflow characterized by the production
of about 20 neutrons per heavy nucleus at temperatures lower than $1.5\cdot 10^9$K. Apart from having a smaller
radius at low temperatures this trajectory is identical to that represented by the first two
columns of this table.}

\end{deluxetable}

%\clearpage
%\begin{deluxetable}{cccc}
%\tabletypesize{\scriptsize}
%\tablecaption{Overprodution factors characterizing nucleosynthesis 
%in the outflow from the 2D simulations that had the highest entropy 
%and electron fraction ($s/k_b\approx 77$, $Y_e =0.57$). 
%\label{nominalWind}}
%\tablewidth{0pt}
%\tablehead{
%\colhead{isotope}
%& \colhead{$\log(X/X_{\odot})$}} 
%\startdata

%\enddata
%\end{deluxetable}

%\clearpage
%\begin{deluxetable}{cccc}
%\tabletypesize{\scriptsize}
%\tablecaption{Overprodution factors characterizing nucleosynthesis in an
%outflow with $s/k_b\approx 150$ and $Y_e=0.57$. 
%\label{s150ye57}}
%\tablewidth{0pt}
%\tablehead{
%\colhead{isotope}
%& \colhead{$\log(X/X_{\odot})$}} 
%\startdata
%\enddata
%\end{deluxetable}

%\clearpage
%\begin{deluxetable}{cccc}
%\tabletypesize{\scriptsize}
%\tablecaption{Overprodution factors characterizing nucleosynthesis in an
%outflow with $s/k_b\approx 190$ and $Y_e=0.65$. 
%\label{25sWind}}
%\tablewidth{0pt}
%\tablehead{
%\colhead{isotope}
%& \colhead{$\log(X/X_{\odot})$}} 
%\startdata
%\enddata
%\end{deluxetable}

\end{document}